\documentclass{ws-procs961x669}            % default, citations in superscript
\usepackage{amsmath}
\usepackage{mathtools}
\usepackage{siunitx}
\begin{document}
\title{Searching for Intermediate Mass Black Holes \\ in the Milky Way's galactic halo}

\author{A. Franco$^*$, A.A. Nucita, F. De Paolis, F. Strafella and M. Maiorano}
\address{Department of Mathematics and Physics ``Ennio De Giorgi'', \\ 
        University of Salento, Via per Arnesano, I-73100, Italy,    \\
	    INFN, Sezione di Lecce, Via per Arnesano, I-73100, Italy,   \\
    	$^*$E-mail: antonio.franco@le.infn.it  }

\begin{abstract}
Intermediate Mass Black Holes (IMBHs) are a class of black holes with masses in the range $10^2 \div 10^5$ $M_\odot$, which can not directly derive from stellar evolution. Looking for these objects and estimating their abundance is important not only for a deeper understanding of their origin but also for unveiling the nature and distribution of the dark matter in the galactic halo.
Since February 2018 to January 2020, the Large and Small Magellanic Cloud have been intensively monitored by the DECAM instrument, installed on the 4m V. Blanco Telescope (CTIO, Chile) with the main objective to discover microlensing events possibly due to IMBHs. 
\newline
Here we outline the developed data analysis pipeline. We have tested it versus known variable sources finding many not previously known variables objects. A few sources show a light curve similar to that expected for a microlensing event, but further analysis is required to confirm the microlensing nature of these events.

For these sources, and in particular for the uncatalogued variable stars, we try to determine if they are periodic or not via a periodogram analysis.
\end{abstract}


\bodymatter

\section{Introduction}\label{hist_notes}

During the last years of XVIII century, the English pastor John Michell and the mathematician Pierre Simon de Laplace imagined the existence of some dark objects, which were called \textit{dark stars}\cite{bhs}. The given name derived from the fact that if the radius of an object with mass $M$ turns out to be $ R=2GM/c^2 $ (where $G$ is Newton's gravitational constant and $c$ is the light speed), it cannot be visible from outside and therefore it would ``appear'' as \textit{dark}.

The first actual mathematical description was provided in 1916 by Karl Schwarzschild with his solution of Einstein's field equations, representing the gravitational field around a non-rotating, spherical, and electrically neutral mass.

Half a century later, in 1967, John Wheeler coined the term \textit{black hole} (BH) to indicate an object whose gravitational field is so large that does not let anything, even light, escape from its gravitational attraction. Unlike the classical formalization, General Relativity predicts the existence of an event horizon located at $ r = r_S = 2GM /c^2$, which is called the Schwarzschild radius.

\section{Black Holes Classification}

There are three main kinds of black holes: stellar, intermediate, and supermassive BHs. 
The most common ones  are stellar and supermassive BHs: the first one, with a mass less than $30 \> M_\odot$, are produced by stellar death, when very massive stars with $M > 25-30\> M_{\odot}$ reach the and of their lives. \newline
Supermassive BHs, with $M \simeq 10^5\div10^{10}\>M_\odot$, are located at the center of almost every galaxy in the Universe and are also responsible for the AGN activity. \newline
The most unknown and mysterious intermediate-mass black holes (IMBHs) have a hypothetical mass in the range $30\div10^5\>M_\odot$ and might provide the Galactic Halo’s dark matter contribution. Their formation and evolution are not well known as we are not able to observe many of these objects. There are four main possible mechanisms for IMBHs \footnote{Searches for IMBHs have been attempted recently in dwarf spheroidal galaxies (dSph) where these objects are expected to form (see, e.g., Ref.\citenum{manni, nucita1, nucita2, nucita3, nucita4, nucita5, reines, lemons}). In fact, extrapolating the fundamental $M_{BH}$ - $M_{Bulge}$ relation (see, e.g., Ref.\citenum{magorrian}, for the supermassive BHs case) down to typical dSph masses, black holes in the typical IMBH mass range arise.} formation \cite{imbh_paper}:
\begin{itemize}
\item Evolution of Population III stars
\item Stellar collisions in dense star clusters
\item Primordial Black Holes \cite{zeld, haw2, haw, frampton, garbel, khlopov}
\end{itemize}

\section{Gravitational microlensing}

Gravitational microlensing may offer an efficient way to detect BHs if they populate the Galactic Halo. This phenomenon shows up when an object (the lens) stands in the way between the observer and a faraway source. Photons coming from the source are deflected by the gravitational field of the lens (see Ref.\citenum{depaolis} for a review). Depending on the phenomenon geometry, we might observe virtual images of the source. If the images are separated by large enough angles, as happens in the case of lensed quasars, one can detect double or multiple images of the same source. In the case of unresolved images, we have a microlensing event. It might be possible to observe the luminosity variation of the source, that is called \textit{amplification}, which follows the \textit{Paczy\'nski law}\cite{pacz1}:
\begin{equation}
    \mu(t)={{u(t)^2+2}\over{u(t)\sqrt{u(t)^2+4}}}
\end{equation}
where $ u(t) = \beta(t)/\theta_E$ is the Lens-Source angular separation ($\beta(t)$) in Einstein’s angle ($\theta_E$) unit.

\section{Searching for Intermediate Mass Black Holes with DECAM}

Searching for IMBHs might be important to obtain significant information regarding the evolution of black holes and constrain their contribution to the halo dark matter. This study could give major contributions about PBHs as dark matter candidates \cite{afshordi, carrkuh}.  This study can be carried out using the microlensing phenomenon and monitoring some regions in the sky with a very high stellar density, such as the Magellanic Clouds.
\newline
The microlensing rate toward the Large Magellanic Cloud (LMC in the following) is estimated to be \cite{griest}:
\begin{equation}
    \Gamma\approx\frac{1.66\times10^{-6}}{(M/M_\odot)^{1/2}} \> \text{event/year}
\end{equation}
and, given $N_0 \approx 10^7$ the number of LMC stars, the number of expected microlensing events for solar mass lenses is :
\begin{equation}
    N_{ev}=N_0\>\Gamma(\text{M})\> t_{exp}\> \varepsilon(\text{M}) \approx 50\> \text{events}
\end{equation}
where $t_{exp}\simeq 2\>\text{years}$ is the duration of the survey and $\varepsilon\approx 0.1$ is the detection efficiency. 
Taking into account microlensing events with a duration up to two years, is possible to estimate the detectable mass limit that is $\lesssim 300 \>M_\odot$: we can therefore observe lenses in the IMBH mass range. 

\subsection{DECAM instrument}

DECAM \cite{decam_book} (Dark Energy CAMera) is a high-performance, wide-field CCD imager mounted at the prime focus of the Victor M. Blanco 4m Telescope (CTIO, Chile). This camera consists of a grid of 62 CCDs (2048$\times$4096 Pixels) which cover a field of view of $\sim 3\> \text{deg}^2$ ($\sim 2.2$ degree wide) with 0.263 arcsecond/pixel resolution. It uses a set of filters similar to those employed in the Sloan Digital Sky Survey (g,r,i,z,Y filters). It is active since September 2012 on behalf of the Dark Energy Survey (DES), finished in January 2019.

\subsection{Blanco/DECAM survey}

The proposal considered for this work, whose Principal Investigator is William Dawson, is related to the current rate of LIGO events that favors such a large abundance that IMBHs would make up the majority of dark matter. A Blanco/DECAM microlensing survey is useful to search for similar IMBHs in our Milky Way\cite{decam_prop}. The survey was designed to intensively observe the Magellanic Clouds for two years, from February 2018 to January 2020, acquiring several images toward 23 LMC fields and 6 SMC fields. In particular, the LMC survey, currently under study, has continuously monitored a portion of sky with $4\text{h}\>40\text{m}<\alpha < 6\text{h}\>20\text{m}$ and $\ang{-73}<\delta<\ang{-65}$. The images have been acquired with a sampling of 7-10 days between each observation night, and a sub-sampling of few hours, acquiring 3-4 images during the same night. The final result is about 40 images acquired for each field and for each filter used (mainly in \textit{g} and \textit{r} bands, with just few images obtained in the \textit{i} band).

In particular, for our aim, we have used 36 resampled images (bias, flat and dark subtracted) in the \textit{g} band acquired toward the Large Magellanic Cloud. 

\subsection{Detection Method}
\def\figsubcap#1{\par\noindent\centering\footnotesize(#1)}
\begin{figure}[h]%
\begin{center}
  \parbox{3in}{\includegraphics[width=3in]{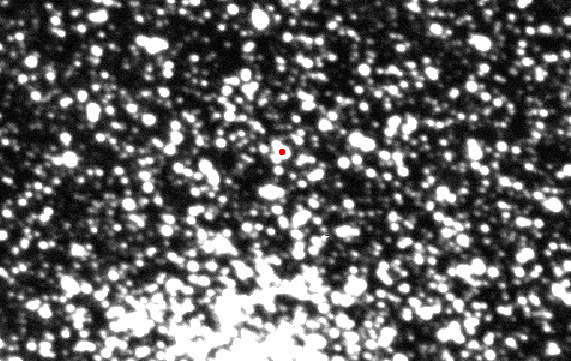}\figsubcap{a}}\\
  %\hspace*{4pt}
  \vspace*{2pt}
  \parbox{3in}{\includegraphics[width=3in]{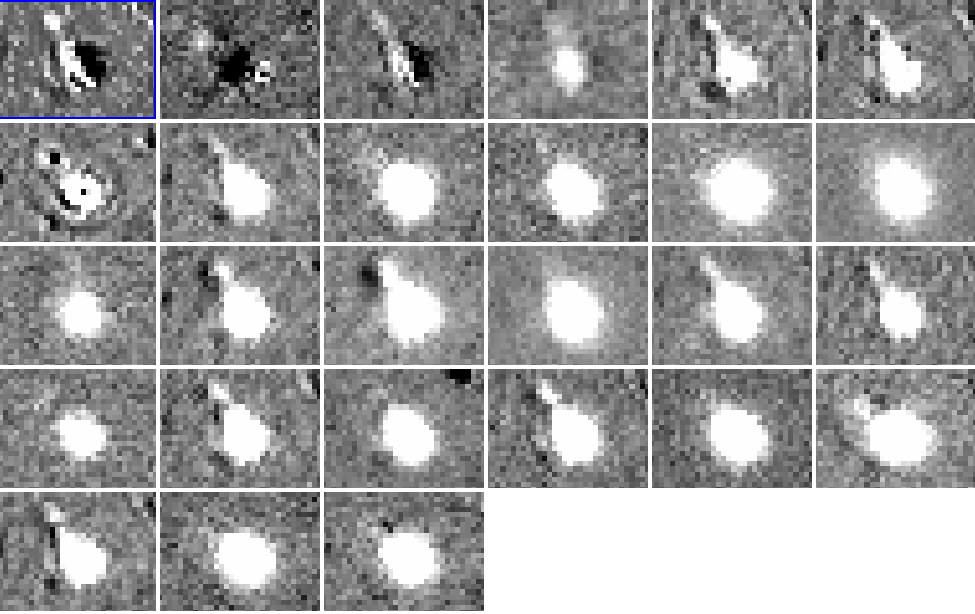}\figsubcap{b}}
  \caption{(a) Small frame within the observed field of view toward LMC. The red dot indicates an identified microlensing candidate at coordinates $\alpha=\ang{84.126738}, \delta=\ang{-68.797069}$. (b) Differential luminosity over time for the microlensing candidate event.}%
  \label{fig::source_candidate}
\end{center}
\end{figure}
In order to search for microlensing events, it is essential the detection of variable sources in the images. This goal can be achieved using the ISIS 2.2 subtraction package \cite{alard1, alard2}. The ISIS software, developed by Christophe Alard since the last years of the 90s, allows to perform the photometry of different objects through images difference, especially in very crowded field like LMC and SMC. 
The software works in multiple steps:
\begin{itemize}
    \item Interpolation: images alignment; 
    \item Creation of a reference image (from N selected ``good'' images from the hole sample) to compute the subtraction process;
    \item Subtraction between the reference image and each image of the sample, highlighting the brightness difference for variables;
    \item Detection of variable sources;
    \item Execution of the photometry for these objects.
\end{itemize}

Once a variable and potentially interesting source is identified, we use the Schwarzenberg-Czerny algorithm \cite{czerny}, providing the phased light curve, and attempting an estimate of the period of the source. This algorithm gives more reliable results, in particular for high signal/noise ratio values, where the Lomb-Scargle algorithm is not efficient enough \cite{czerny2}.

\subsection{Microlensing candidate analysis}

\begin{figure}[h]
\begin{center}
\includegraphics[width=3in]{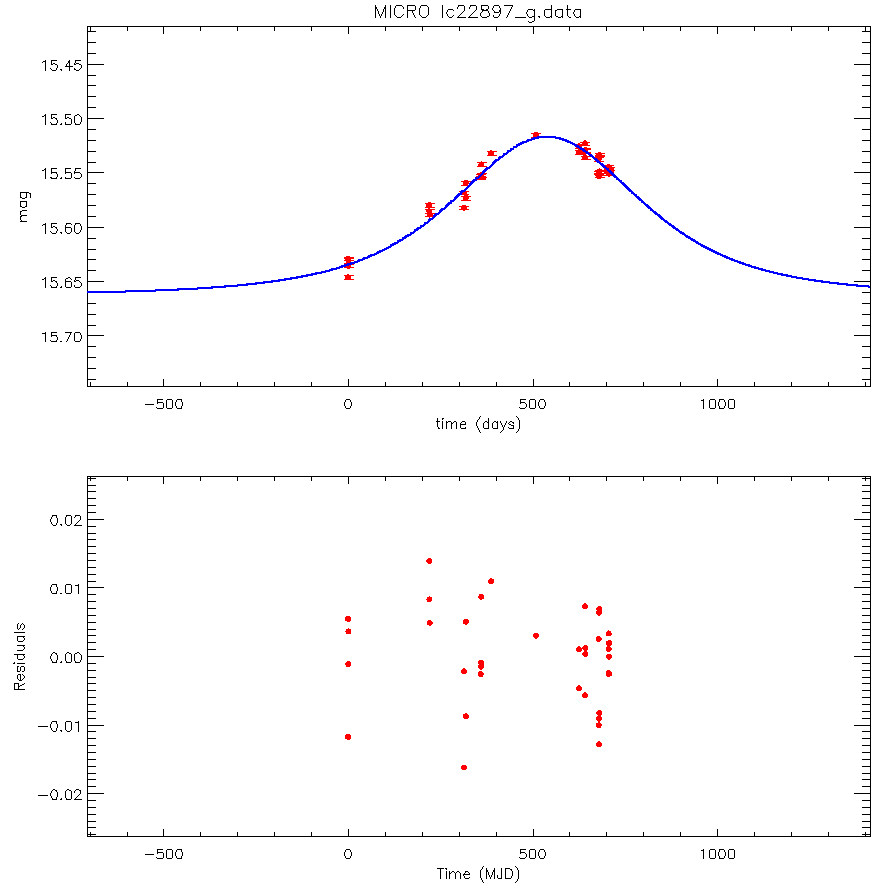}
\end{center}
\caption{The upper box shows the magnitude of the variable sources over two years of data (red dots) and the fitted Paczy\'nski curve. The bottom box shows the residuals.}
\label{fig::fit}
\end{figure}
\begin{figure}[h]%
\begin{center}
  \parbox{2.1in}{\includegraphics[width=2.1in]{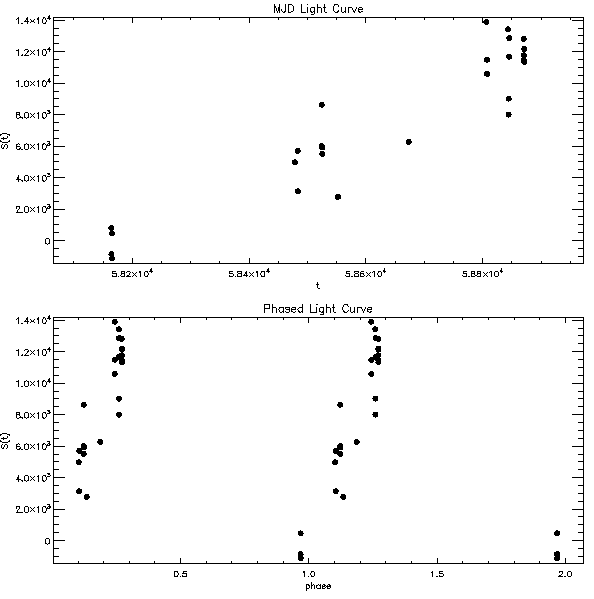}\figsubcap{a}}
  \hspace*{4pt}
  \vspace*{4pt}
  \parbox{2.1in}{\includegraphics[width=2.1in]{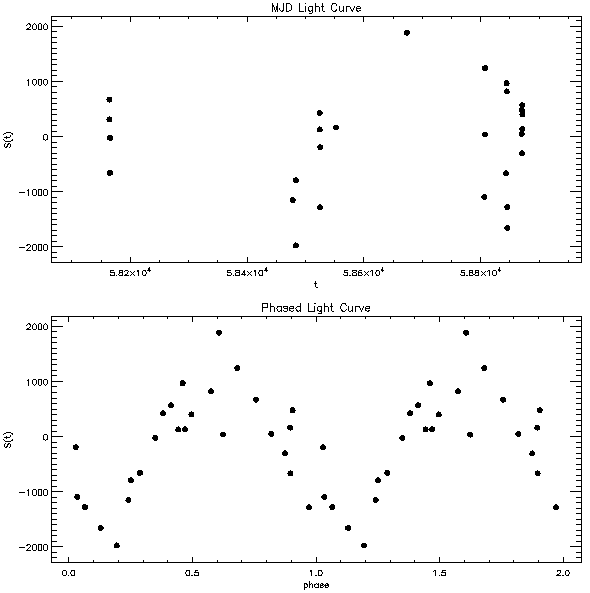}\figsubcap{b}}
  \parbox{2.1in}{\includegraphics[width=2.1in]{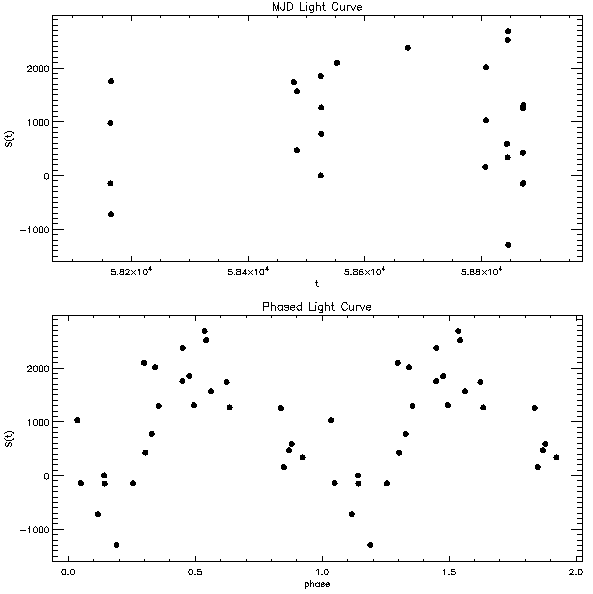}\figsubcap{c}}
  \caption{The MJD light curves (upper panel) and phased light curve (lower panel) for three different variable sources. (a) Non-periodic variable and (b), (c) eclipsing binary candidates.}%
  \label{fig::periodics}
\end{center}
\end{figure}
One of the most interesting variable sources detected so far is a microlensing like curve that seems to be associated with an object in the direction of LMC ($\alpha=\ang{84.126738}, \delta=\ang{-68.797069}$). The field of view  containing this source is shown in \fref{fig::source_candidate}(a). 

Using the ISIS software it appears that this source is clearly variable, as shown in \fref{fig::source_candidate}(b). The corresponding light curve is shown in \fref{fig::fit} from which it is apparent a Paczynski-like behavior, typical of a microlensing event. The fit and the residuals are shown in the upper and lower panels of \fref{fig::fit}, respectively.

The fit returns an estimate of the event Einstein time $t_E=(246.4 \pm 31.7) \> \text{days}$ and the baseline magnitude $m_0=15.66 \pm 0.01$. We note that the indicated magnitude represents the instrumental magnitude of the camera, without any photometric correction. In addition, it is important to understand the geometry of the phenomenon in order to estimate the lens mass. For these reasons new images toward the particularly interesting fields will be acquired in the next future in order to provide a better and complete results, that will be published after completion of the work.

\subsection{Periodic variable sources candidate analysis}

ISIS 2.2 and the Schwarzenberg-Czerny algorithm enable the recognition of a few periodic variables and allow to obtain an estimate of their period. This can be used both to confirm previously observed sources and to detect new ones. The observed data for three variable sources toward the LMC are shown in \fref{fig::periodics}. Panel (a) shows the light curve for non-periodic variable source, at coordinates $\alpha_a=\ang{84.530329}, \delta_a=\ang{-68.791993}$, displaying the increasing flux over time and highlighting the absence of a particular periodicity in the phased light curve. Panels (b) and (c) show two sources, at coordinates $\alpha_b=\ang{84.486772}, \delta_b=\ang{-68.799780}$ and $\alpha_c=\ang{84.303917}, \delta_c=\ang{-68.741760}$, displaying the variation of the flux with respect to the time (upper) and over two phases (bottom), enabling the estimate of the period that is $P_b\simeq1.69$ days and $P_c\simeq0.31$ days.

\section{Conclusions}

The detection of periodic and non-periodic variable sources has allowed confirming the goodness of the developed pipeline. In particular, the two years time window, the sampling and the efficiency of the DECAM instrument have shown the possibility to detect microlensing events. So far, many known variable sources have been identified in four DECAM fields of view - toward LMC - and some unknown variables are waiting to be studied more deeply. A few possible microlensing event candidates (e.g. that shown in \fref{fig::fit}), which wait for confirmation, have been also found in the data. 

The contents presented in this paper  will be discussed in more details in the  the paper ``\textit{Searching for galactic halo Intermediate Mass Black Holes through gravitational microlensing}'', by A. Franco et al., currently under preparation. 

\section*{Acknowledgments}

We acknowledge the support by the Euclid and TAsP (Theoretical Astroparticle Physics) projects of the Istituto Nazionale di Fisica Nucleare (INFN).

%\bibliography{ws-pro-sample}

%\newpage

\end{document}